# One-dimensional ferromagnetic edge contacts to two-dimensional graphene/h-BN heterostructures

Bogdan Karpiak,[1] André Dankert,[1] Aron W. Cummings,[2] Stephen R. Power,[2] Stephan Roche[2,3] and Saroj P. Dash[1*]

[1] *Department of Microtechnology and Nanoscience, Chalmers University of Technology, SE-41296, Göteborg, Sweden*

[2] *Catalan Institute of Nanoscience and Nanotechnology (ICN2), CSIC and The Barcelona Institute of Science and Technology, Campus UAB, Bellaterra, 08193 Barcelona, Spain*

[3] *ICREA—Institució Catalana de Recerca i Estudis Avançats, 08010 Barcelona, Spain*

\* Email: saroj.dash@chalmers.se

## Abstract

We report the fabrication of one-dimensional (1D) ferromagnetic edge contacts to two-dimensional (2D) graphene/h-BN heterostructures. While aiming to study spin injection/detection with 1D edge contacts, a spurious magnetoresistance signal was observed, which is found to originate from the local Hall effect in graphene due to fringe fields from ferromagnetic edge contacts and in the presence of charge current spreading in the nonlocal measurement configuration. Such behavior has been confirmed by the absence of a Hanle signal and gate-dependent magnetoresistance measurements that reveal a change in sign of the signal for the electron- and hole-doped regimes, which is in contrast to the expected behavior of the spin signal. Calculations show that the contact-induced fringe fields are typically on the order of hundreds of mT, but can be reduced below 100 mT with careful optimization of the contact geometry. There may be additional contribution from magnetoresistance effects due to tunneling anisotropy in the contacts, which need to be further investigated. These studies are useful for optimization of spin injection and detection in 2D material heterostructures through 1D edge contacts.

## Introduction

Graphene has been shown to be a promising material for spin-polarized electron transport due to its low spin-orbit coupling and negligible hyperfine interactions[1–3]. Recently, long-distance spin transport and electrical control over spin signal and lifetimes have been achieved in graphene at room temperature[1,2,4–8]. These experiments have shown that the performance of graphene spintronic devices is significantly affected by the quality of the contacts to the graphene channel[1]. Conventional spin transport experiments use top ferromagnetic metal/tunnel barrier contacts of micrometer width on graphene channels for the purpose of spin injection and detection[1–3]. However, the fabrication of atomically smooth oxide tunnel barriers is challenging, as they usually contain pinholes, roughness, and defects[9–11]. The use of such wide contacts is known to limit the device performance due to inhomogeneous injection and detection of spins, and due to interface-induced spin dephasing under the contacts[9–11]. Additionally, at the nanoscale graphene edges can become important and are predicted to host spin-polarized edge states, giving rise to spin filtering[12,13]. Such spintronic properties are also predicted to be tunable by external electric fields[12,13]. If realized, this would add the spin degree of freedom to graphene-based devices and circuits, where spin currents can be generated and injected from zigzag nanoribbons to graphene without the need of ferromagnetic spin injectors.

Here we present the fabrication and characterization of a device that utilizes 1D ferromagnetic edge contacts to graphene/hexagonal boron nitride (h-BN) heterostructures. The non-local spin-valve measurements are found to show spurious magnetoresistance effects due to local Hall effects. Our detailed measurements in different geometries and gate voltages, as well as calculations show that the ferromagnetic contact-induced fringe fields give rise to such a signal.

## Results and discussion

A schematic representation and optical microscopy picture of the device are shown in Figs. 1a and 1b, respectively. The devices with 1D edge contacts to graphene/h-BN heterostructures were fabricated using the process steps presented in Fig. 1c. The shape anisotropy of the ferromagnetic contacts (different widths) was used to achieve different switching fields. The details about the fabrication steps of heterostructure and 1D edge contacts are described in Methods section.

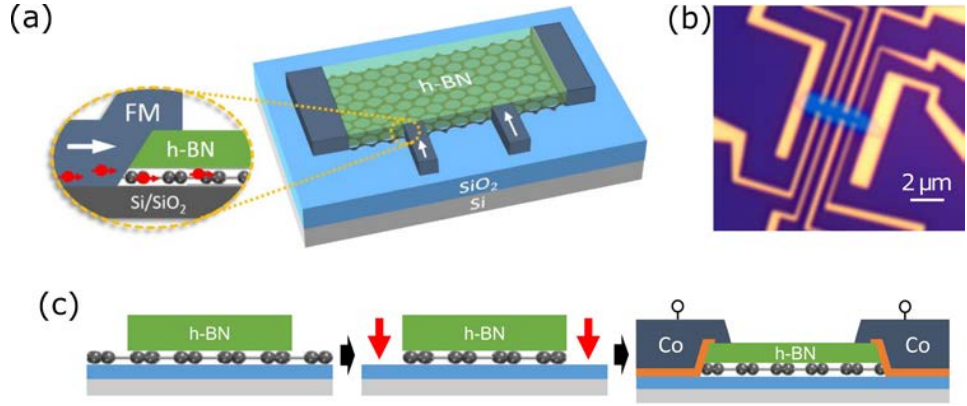

*Figure 1:* *Fabrication of graphene/h-BN heterostructures with 1D ferromagnetic (TiO$_2$/Co) edge contacts. (a) Schematic representation and (b) optical microscopy picture of the fabricated device. (c) Schematics of the fabrication process steps including preparation of heterostructures, their patterning, and fabrication of contacts.*

The basic characterization of the fabricated graphene/h-BN heterostructure is depicted in Fig. 2. The Raman spectrum of the heterostructure is shown in Fig. 2a, where the G and 2D peaks related to graphene[14] are visible at positions 1568 cm$^{-1}$ and 2632 cm$^{-1}$ respectively, and the h-BN peak[15] is at 1343 cm$^{-1}$. The 1D edge contact resistances to graphene were found to be in the range of 4-16 kΩ. The electrical properties of the graphene channel were characterized by gate voltage ($V_g$) dependence in both the local (Fig. 2b) and nonlocal ($R_{NL} = V_{NL}/I_{bias}$, Fig. 2c) configurations. The charge neutrality (Dirac) point of graphene was found to be in the range of $V_D \sim$ -5 to +26 V for different channels measured in local or nonlocal configurations between different contacts due to variations of doping levels within the graphene sheet. The presence of a measurable nonlocal resistance $R_{NL}$ in the detection circuit indicates a charge current spreading outside the bias current circuit, giving rise to an Ohmic resistance contribution $R_s \cdot e^{-\pi L/W}$, where $R_s$ is graphene sheet resistance, and $L$ and $W$ are the length and width of the graphene channel, respectively[16].

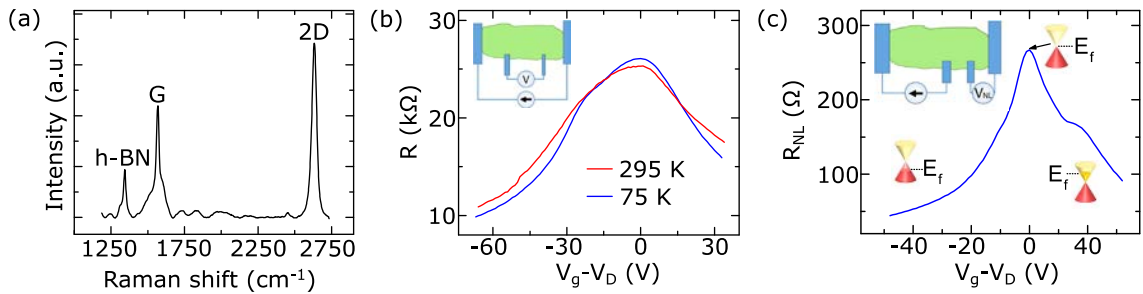

*Figure 2:* *Characterization of graphene/h-BN heterostructures with 1D edge contacts. (a) Raman spectrum of the graphene/h-BN heterostructure. (b) Graphene channel resistance as a function of back gate voltage ($V_g$) at 295 K (red) and 75 K (blue) for a local measurement configuration (inset). (c) Nonlocal channel resistance ($R_{NL} = V_{NL}/I$) as a function of back gate voltage $V_g$ at 75 K for the nonlocal measurement configuration (inset). The horizontal axes in (b) and (c) are plotted as $V_g$-$V_D$.*

Next, nonlocal magnetoresistance measurements were carried out in the 1D edge contact devices as shown in Fig. 3a. We observed a nonlocal voltage $V_{NL}$ with single switching and hysteresis behavior while sweeping the in-plane magnetic field aligned with the contacts at fixed bias currents (Fig. 3b). By changing the current direction, a similar switching signal with opposite sign was observed. The complete bias dependence of the signal is presented in Fig. 3c, which shows a linear dependence in the measured bias range. The temperature dependence of the magnetoresistance signal $V_{NL}$ was measured at a constant current bias of $I = +15$ µA (Fig. 4). We observed a decay of the switching amplitude of $V_{NL}$ ($\Delta V_{\uparrow\downarrow}$) with increasing temperature, which could be measured up to 200 K.

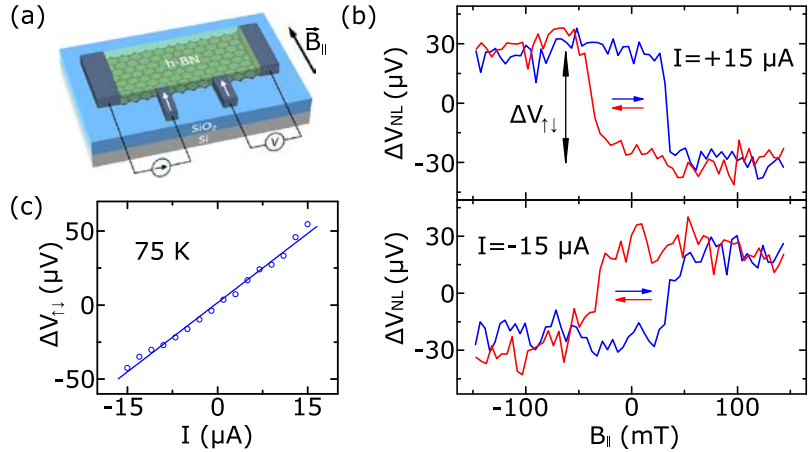

*Figure 3:* *Magnetoresistance measurements and bias dependence of the spin valve device. (a) Schematic of the nonlocal measurement configuration. (b) Measured nonlocal voltage $V_{NL}$ as a function of external in-plane magnetic field ($B_{in}$) at $I=\pm 15$ µA at 75 K. The $B_{in}$ sweep directions are indicated by red and blue arrows. A baseline linear background voltage is subtracted from the measured data. (c) Current bias dependence of the magnetoresistance switching amplitude $\Delta V_{\uparrow\downarrow}$ at 75 K.*

The presence of only one step of magnetoresistance for each sweep direction is not typical for spin signals that arise from spin injection and detection[6,11,17–20]. For comparison, a typical spin-valve signal for top ferromagnetic contacts to graphene is shown in Supplementary Fig. S1b). In the latter case, at least two steps are usually visible for each sweep direction, when both injector and detector contacts switch their magnetization direction[6,11,17–20]. The coercivity values of the ferromagnetic contacts used here are within the typical sweeping field range, which we have also verified with higher field sweep ranges (Supplementary Fig. S2b). A single-step switching signal can arise if the graphene edge itself generates a spin current; however, such effects are only expected in graphene nanoribbons and not in the micrometer-scale devices used here[12,13]. The quantum spin Hall effect is also unlikely to be responsible for the observed signal considering the negligible spin-orbit coupling in graphene. Additional confirmation of the absence of spin transport in our 1D contact devices comes from the out-of-plane field sweeps, where no Hanle spin precession signal is observed (Supplementary Figure S2c). The continuous linear change of magnetoresistance here (inset in Supplementary Fig. S2c) could be due to Hall effect in the presence of continuous sweeps of external perpendicular magnetic field, in contrast to abrupt changes of stray fields as e.g. in Fig. 3. Therefore, the observed single-switching magnetoresistance with 1D ferromagnetic contacts could be due to the local Hall effect in the graphene in the presence of stray magnetic fields emanating from the edges of ferromagnetic contacts[10,21–24].

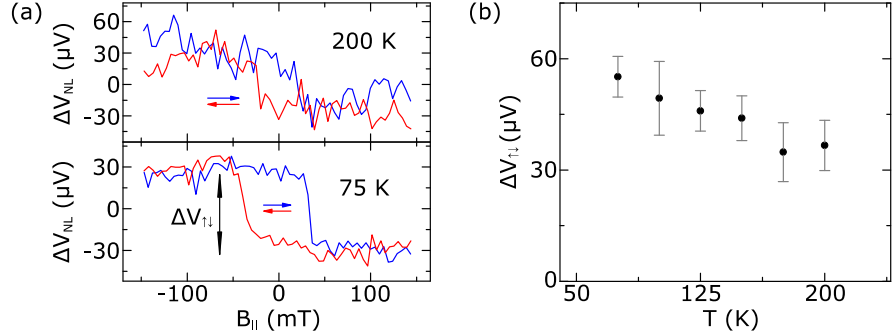

*Figure 4:* *Temperature dependence of the magnetoresistance signal. (a) Nonlocal voltage as a function of applied external in-plane magnetic field at 200 K (top panel) and 75 K (bottom panel) at 15 µA current bias. The field sweep directions are indicated by arrows and a linear background voltage is subtracted from the measured data. (b) Temperature dependence of the switching amplitude (ΔV$_{↑↓}$) at 15 µA current bias.*

In order to further clarify the origin of the magnetoresistance, gate-dependent measurements were carried out, where we tune the concentration and type of the charge carriers in graphene. Figure 5a shows the magnetoresistance switching with the application of $V_g$ = ±30 V, where a change in the sign of the Δ$V_{↑↓}$ is observed due to electron or hole conduction of the graphene channel. The complete gate dependence of Δ$V_{↑↓}$, along with the channel resistance is shown in Fig. 5b, revealing a correlation between the sign of Δ$V_{↑↓}$ and type of charge carrier in graphene. The absolute value of Δ$V_{↑↓}$ is found to have two maxima near the charge neutrality point in graphene, where the charge density $n_{2D}$ is minimal. These results support the argument in favor of a local Hall effect-dominated magnetoresistance switching. At the same time, this gives additional evidence to rule out any spin-related nature of the observed switching, since the spin signal should not change sign with a change of charge carrier type[17].

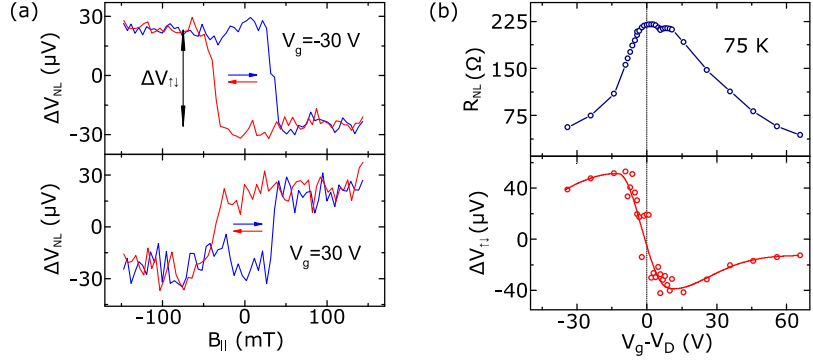

*Figure 5:* Gate dependence of the magnetoresistance signal. (a) $V_{NL}$ as a function of in-plane external magnetic field sweeps at $V_g=-30$ V (top panel) and $V_g=30$ V (bottom panel), measured with $I = 15$ μA at 75 K. A linear baseline offset voltage is subtracted from the raw data. (b) Dependence of the nonlocal graphene resistance $R_{NL} = \Delta V_{NL}/I$ on the back gate voltage $V_g-V_D$ (top panel) and the nonlocal voltage step $\Delta V_{\uparrow\downarrow}$ (bottom panel). $V_D$ is the Dirac point of graphene.

Next, regular Hall measurements were performed (Fig. 6a) on the same device at 75 K to correlate with the magnetoresistance measurements. Figure 6b shows a change in the sign of the slope of the Hall response $V_H$ in the graphene channel by the application of a gate voltage $V_g= \pm40$ V, due to a change in conduction from electrons to holes. The full gate-dependent Hall measurements were performed by sweeping $V_g$ from electron to hole conduction across the Dirac point in the presence of different perpendicular magnetic fields (Fig. 6c). We observe a change in the amplitude and sign of $V_H$ with the applied $V_g$. A clear similarity is also observed between the $V_g$ dependence of the magnetoresistance signal $\Delta V_{\uparrow\downarrow}$ (Fig. 5b) and the regular Hall voltage $V_H$ (Fig. 6c). This similarity is justified due to the similar origin of the local Hall magnetoresistance and the regular Hall effects, which are due to stray magnetic fields from the ferromagnetic contacts, or the Lorentz force acting on moving charges in the presence of a perpendicular external magnetic field, respectively.

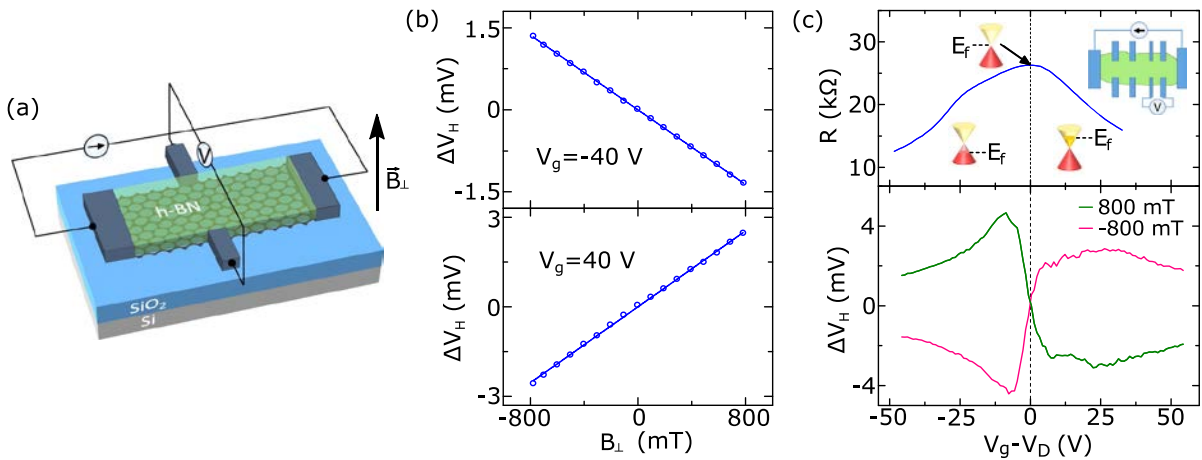

*Figure 6:* Hall effect in graphene. (a) Schematic of the Hall measurement configuration. (b) Hall voltage $V_H$ as a function of the perpendicular magnetic field, measured with $V_g = \pm40$ V at 75 K. (c) Gate dependence of the *graphene resistance (top) and Hall voltage $V_H$ (bottom) at 75 K for different perpendicular magnetic fields. The reference signal measured at a magnetic field of 0 T is subtracted from the Hall response at other magnetic fields. Measurements at room temperature for this device are reported elsewhere[25].*

In order to quantify the impact of the local Hall effect on the magnetoresistance signal, we use the Biot-Savart law to numerically calculate the stray magnetic fields that can arise from the 1D edge contacts. The contact geometry we consider is shown in Fig. 7a. To describe the contact we assume a surface current density given by the saturation magnetization of cobalt, $M_s$ = 1.42 MA/m, which is parallel to the +x (-x) axis on the top (bottom) surfaces of the side view, and parallel to the +z (-z) axis on the left (right) surfaces of the top view. Away from the sloped region, this magnetizes the cobalt along the y-axis. In the sloped region, this surface current also yields an out-of-plane component to the polarization. For metal surfaces parallel to the principal axes, the stray magnetic field can be calculated analytically[26], while the sloped portion of the contact requires a numerical solution.

In Fig. 7b we show the spatial distribution of the perpendicular component of the stray magnetic field $B_z$ within the graphene layer, using the experimental contact geometry (with $t_{h-BN,lower}$ = 0 and the rest of the parameters listed in the caption of Fig. 7). The black dashed lines indicate the contact metal edges. Here we see that the stray field can be quite strong, reaching more than 600 mT at the graphene edge. In addition, the end of the top portion of the ferromagnetic contact also induces a strong perpendicular stray field on the order of 350 mT. Near the 1D edge contact to graphene, the stray field decays from 600 mT to zero over a distance of 50 nm. This average field of 300 mT corresponds to an average magnetic length of ~50 nm, indicating that the stray field can indeed induce a significant Hall effect at the charge injection/detection point.

In Fig. 7c we plot $B_z$ along the middle of the contact, indicated by the white dashed line in Fig. 7b. This plot shows that by increasing the thickness of the bottom h-BN layer, the magnitude of the stray field at the graphene edge can be reduced by nearly a factor of three for a typical set of experimental parameters. This reduction occurs because the stray fields due to the top and bottom surfaces of the contact tend to cancel one another as the vertical position of the graphene layer increases. While the perpendicular stray field at the injection edge is still relatively large, ~250 mT, further optimization of the contact geometry can reduce this. For example, tuning the ratio of $t_{h-BN,upper}/t_{h-BN,lower}$ can reduce $B_z$ at the injection edge (Fig. 7d), but this can result in deeper penetration of stray fields into the graphene sheet, and in sign changes of $B_z$ at different distances from the graphene edge (Fig. 7e). Generally, with thicker top and bottom h-BN layers, the stray field at the edge can be reduced to below 100 mT (Fig. 7f and Supplementary Fig. S3a). A shallower etching angle θ could also reduce the stray fields at the graphene edge (Supplementary Fig. S3b), but this is difficult to tune experimentally. Additionally, thicker top h-BN layer can significantly reduce the stray fields at the contact edge corresponding to y = 300 nm (Supplementary Fig. S3c).

It should also be noted that this calculation overestimates the stray field by a factor $M_s/M_r$, where $M_r$ is the remanent magnetization of the contact.

Similar magnetoresistance switching effect can also arise due to tunneling anisotropic magnetoresistance (TAMR)[27–32]. In magnetic tunnel junctions, TAMR signal strongly depend on the orientation of the magnetization with respect to the current direction, crystallographic axes, spin-orbit interaction (SOI) and density of states (DOS) anisotropies in the materials. In such devices, TAMR signals were measured at low temperatures and the magnitude is found to enhance in presence of heavy elements. However, the TAMR signal used to be generally observed only at very low temperatures below 100 K due to sampling of wider region of the tunneling DOS at higher temperatures. In contrast, our magnetoresistance signal in graphene devices is persistent up to 200 K and vanishes only due to increased noise level of the junctions at higher temperatures. The TAMR effects in literature are also known to have its characteristic bias, temperature and angle dependence, and are strongly dependent on the interface SOI. Further detailed investigations are required to identify the contribution of TAMR in the observed magnetoresistance signal (in addition to the stray Hall effect) in our graphene devices with 1D ferromagnetic contacts.

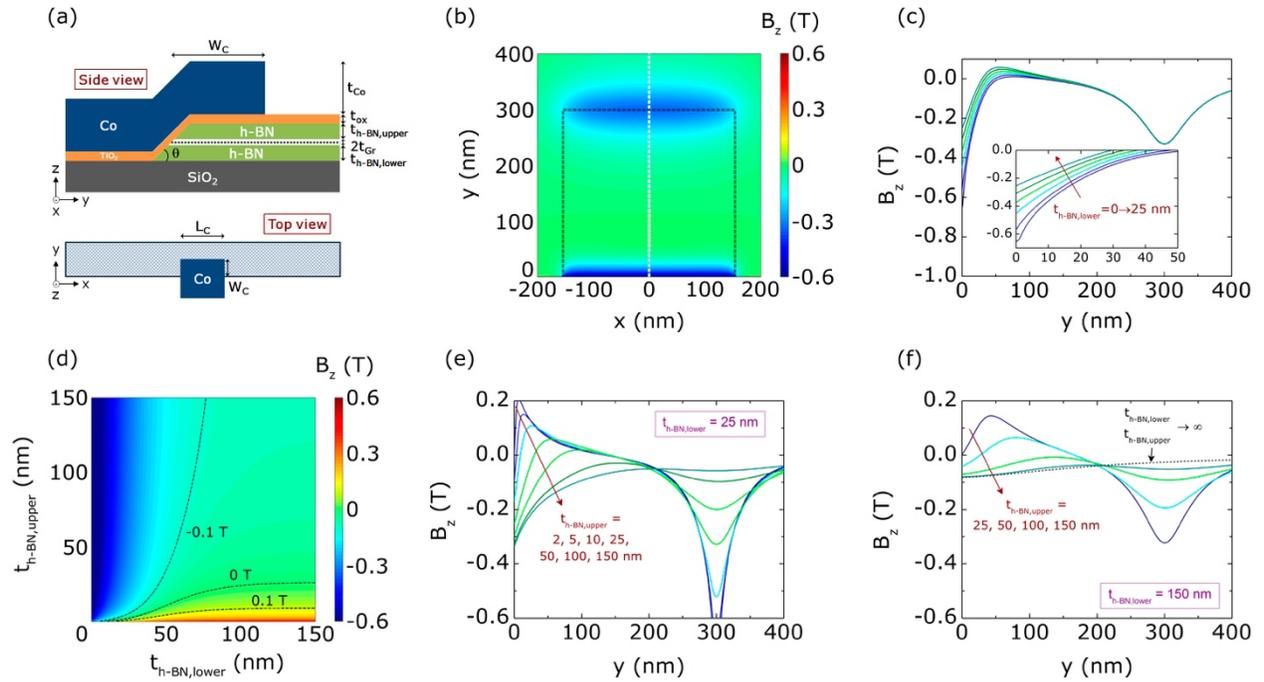

***Figure 7:*** *Stray magnetic fields from 1D ferromagnetic edge contacts. (a) Contact geometry considered for the stray field calculations. (b) Profile of the perpendicular stray field $B_z$ within the graphene layer, assuming $L_C$ = 310 nm, $W_C$ = 300 nm, $t_{Co}$ = 65 nm, $t_{ox}$ = 1 nm, $t_{h-BN,upper}$ = 25 nm, $t_{Gr}$ = 0.35 nm, $t_{h-BN,lower}$ = 0, and $\theta$ = 45°. (c) Stray magnetic field along the center of the contact (x = 0) for different thicknesses of $t_{h-BN,lower}$, where the y-position is relative to the graphene edge. (d) Profile of $B_z$ at different thicknesses of $t_{h-BN,lower}$ and $t_{h-BN,upper}$ at the graphene edge (x = y = 0). (e) $B_z$ as a function of distance from the graphene edge for $t_{h-BN,lower}$ = 25 nm and (f) 150 nm at different $t_{h-BN,upper}$.*

## Conclusions

We have investigated the operation of a graphene device with 1D ferromagnetic edge contacts in a nonlocal configuration. The analysis of the measurements indicates a local Hall effect-dominated phenomenon in the presence of charge current spreading outside the injection circuit, resulting in a single-step magnetoresistance switching. This local Hall magnetoresistance can arise due to the Lorentz force acting on moving charges in the presence of ferromagnetic stray fields from the 1D edge contacts. This origin of the signal is also supported by the absence of a Hanle signal and the change in sign of the magnetoresistance with the change in carrier type from electron to hole in the graphene transport channel. Our calculations reveal that with proper optimization of the contact geometry, the magnitude of the perpendicular stray field can be reduced to below 100 mT at the contact edge. Further investigations are required to identify other contributions to the measured signal such as TAMR in our graphene devices. These findings provide insight into the phenomena that hinder the operation of nonlocal spin valve devices with 1D edge contacts, and they lay the foundation for further developments.

**Methods**

To fabricate h-BN encapsulated graphene devices, first the CVD graphene was wet-transferred on Si/SiO$_2$ substrate (from Graphenea). After cleaning and annealing in Ar/H$_2$ atmosphere, h-BN flakes were dry-transferred using the scotch-tape method. The graphene was patterned by means oxygen plasma etching where h-BN flakes were used as a mask. Electron beam lithography and electron beam evaporation of metals were utilized to fabricate the contacts (1 nm TiO$_2$/65 nm Co), which formed 1D connection to encapsulated graphene. Electronic and magnetotransport measurements were performed in a cryostat by means of Keithley 6221 current source and Keithley 2182A nanovoltmeter. Gate voltage was applied using Keithley 2612 source metre.

**Acknowledgements**

BK, AD and SPD acknowledge financial supports from EU Graphene Flagship (No. 604391), EU FlagEra project (No. 2015-06813), Swedish Research Council grants (No. 2012-04604 and No. 2016-03658), Graphene center and AoA Nano program at Chalmers University of Technology. BK acknowledges scholarship from EU Erasmus Mundus Master of Nanoscience and Nanotechnology. SRP acknowledges funding from the European Unions Horizon 2020 research and innovation programme under the Marie Sklodowska-Curie grant agreement No 665919. SR and AWC acknowledge the Severo Ochoa Program (MINECO, Grant SEV-2013-0295), the Spanish Ministry of Economy and Competitiveness (MAT2012-33911), and Secretaría de Universidades e Investigación del Departamento de Economía y Conocimiento de la Generalidad de Cataluña.

*Phys. Rev. Lett.* **100** 87204.

# Supplementary materials

# One-dimensional ferromagnetic edge contacts to two-dimensional graphene/h-BN heterostructures

Bogdan Karpiak,[1] André Dankert,[1] Aron W. Cummings,[2] Stephen R. Power,[2] Stephan Roche,[2,3] Saroj P. Dash[1*]

[1] Department of Microtechnology and Nanoscience, Chalmers University of Technology, SE-41296, Göteborg, Sweden.
[2] Catalan Institute of Nanoscience and Nanotechnology (ICN2), CSIC and The Barcelona Institute of Science and Technology, Campus UAB, Bellaterra, 08193 Barcelona, Spain.
[3] ICREA—Institució Catalana de Recerca i Estudis Avançats, 08010 Barcelona, Spain

## S1. Spin transport in nonlocal measurement configuration with top ferromagnetic contacts.

To show the typical magnetoresistance signal that originates from spin transport in graphene, a device utilizing $TiO_2$/Co top contacts is measured in the nonlocal configuration (Fig. S1a). Figure 1b shows the spin-valve measurement by sweeping the in-plane magnetic field, which reveals two-step magnetoresistance switching in each sweep direction. Figure S1c shows a typical Hanle spin precession signal with out-of-plane magnetic field sweep.

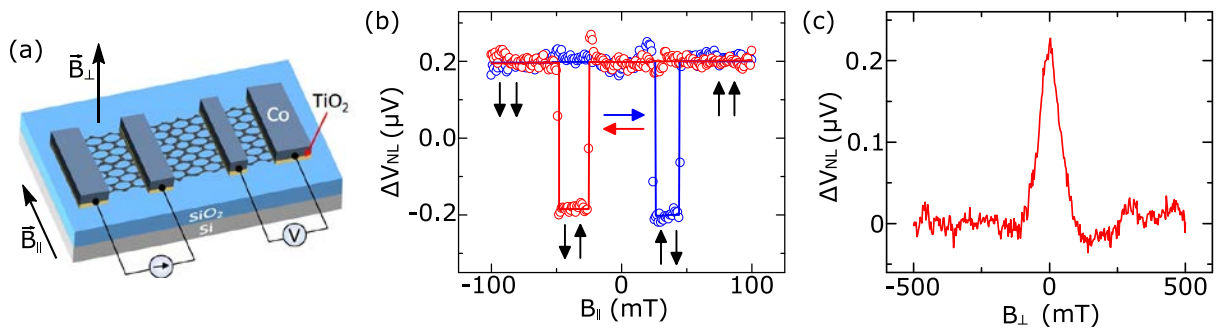

**Figure S1:** *Spin transport measurements in nonlocal configuration with top ferromagnetic contacts. (a) Device schematic and measurement configuration. (b) Measured nonlocal voltage $V_{NL}$ as a function of external in-plane field ($B_∥$) sweep. The magnetic field sweep directions are indicated by the red and blue arrows. Black arrows indicate alignment of ferromagnetic contacts. (c) Hanle spin precession signal $V_{NL}$ as a function of out-of-plane magnetic field ($B_⊥$) sweep. Measurements were performed at an applied current bias I=10 µA. A baseline linear background voltage is subtracted from the measured data.*

## S2. Control experiments to rule out the spin nature of the observed switching phenomenon in 1D contact geometry.

For 1D ferromagnetic contact configuration (Fig. S2a), the gate-dependent measurements presented in main manuscript (Fig. 5b) allowed us to eliminate the spin origin of the observed signal. However, to further prove that the measured one-step magnetoresistance signal does not have additional switching steps, we measured the signal up to higher magnetic fields. No additional steps of magnetoresistance were observed at fields above $\pm 300$ mT as shown in Fig. S2b, which is larger than the coercivity of ferromagnetic contacts used (Fig. S1b). Additional confirmation of the absence of spin transport in our measurements also comes from the absence of Hanle spin precession signal (main panel in Fig. S2c). In the latter case, no step-like magnetoresistance signal is expected since no abrupt change in the effective perpendicular magnetic field is present. However, continuous linear change of magnetoresistance can be observed (inset in Fig. S2c), which could be due to Hall effect in the presence of continuous sweeps of external perpendicular magnetic field, in contrast to abrupt changes of stray fields as e.g. in Fig. 3.

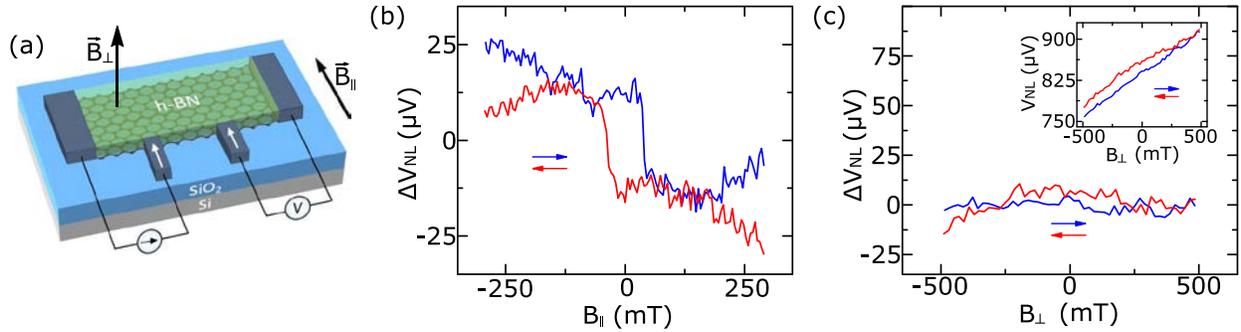

*Figure S2:* Control experiments in 1D ferromagnetic contact geometry. (a) Schematic of the nonlocal measurement configuration. (b) Measured nonlocal voltage $V_{NL}$ as a function of external in-plane magnetic field ($B_\parallel$) with a wide range of field sweep at $I=10$ μA at 75 K. (c) $V_{NL}$ as a function of external out-of-plane magnetic field ($B_\perp$) with subtracted baseline linear background voltage (main panel) and raw data (inset). Measurements were performed at an applied current bias $I=10$ μA at 75 K. The magnetic field sweep directions are indicated by the red and blue arrows.

## S3. Further optimization of contact geometry for the reduction of stray fields.

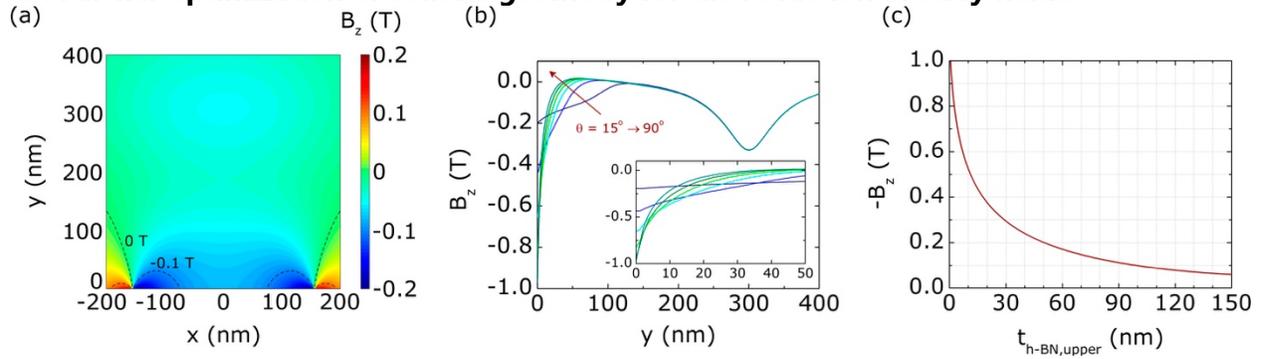

*Figure S3:* Stray magnetic fields from 1D ferromagnetic edge contacts. (a) Profile of the perpendicular stray field $B_z$ within the graphene layer, assuming the contact geometry and parameters in Fig. 7 with $t_{h-BN,upper}$ = $t_{h-BN,lower}$ = 150 nm. (b) Stray magnetic field profile (x = 0) for different etching angles θ of the 1D edge contact for $t_{h-BN,upper}$ = 25 nm and $t_{h-BN,lower}$ = 0. A shallower angle reduces stray fields at the edge, but extends nonzero stray fields further into the graphene. (c) Stray magnetic field in graphene below the edge of the FM contact (x = 0, y = 300 nm) for different thicknesses of the top h-BN layer. A thicker top h-BN layer reduces the stray field at the edge of the FM contact.